\documentclass[onecolumn]{aastex63}

\usepackage{float}
\usepackage{amsmath}
\usepackage{afterpage}

\shorttitle{qCMOS Detectors}
\shortauthors{Roth}

\begin{document}

\title{qCMOS detectors and the case of hypothetical primordial black holes in the solar system, near earth objects, transients, and other high cadence observations}

\correspondingauthor{Martin M. Roth}
\email{mmroth@aip.de}

\author[0000-0003-2451-739X] {Martin M. Roth}
\affiliation{Deutsches Zentrum f\"ur Astrophysik (DZA), Postplatz 1, 02826 G\"orlitz, Germany}
\affiliation{Leibniz-Institut für Astrophysik Potsdam (AIP), An der Sternwarte 16, 14482 Potsdam, Germany}
\affiliation{Institut f\"ur Physik und Astronomie, Universit\"at Potsdam, Karl-Liebknecht-Straße 24/25, 14476 Potsdam, Germany}

\begin{abstract}
Recent progress with CMOS detector development has opened  new parameter space for high cadence time resolved imaging of transients and fast proper motion solar system objects. Using computer simulations for a ground-based 1.23~m telescope, this research note illustrates the gain of a new generation of fast readout low noise qCMOS sensors over CCDs and makes the case for high precision monitoring of asteroid orbits that can potentially shed light on the hypothetical existence of low mass primordial black holes, as well as for other applications requiring high speed imaging.
\end{abstract}


\section{Introduction}
\label{sec:intro}
Over the past decade, the improvement of CMOS image sensors has made this technology gradually more attractive for astronomical instrumentation. Although currently parameters like pixel size, overall sensor area, and quantum efficiency at longer wavelengths still favor conventional CCDs over CMOS, the development of low noise and fast readout qCMOS sensors \citep{Ma+2022} holds the promise of a paradigm change. \citet{Roberts+2024} have evaluated the commercially available Hamamatsu ORCA-Quest qCMOS camera for measurements of quantum spatial correlations and come to conclusions that encourage the use of such a camera also for astronomical imaging. This research note motivates the application of fast cadence qCMOS imaging, remembering that time domain is a priority science area of the Decadal Survey Astro2020, for observations requiring a time resolution of order 1~s, e.g., exoplanet transits of M stars that are affected by flares \citep{Tovar-Mondoza+2022}, near Earth objects, space debris, precision tracking of asteroids in the search for hypothetical primordial black holes within the solar system \citep{Tran+2024,Thoss+2024}, or fast transients of any kind. The advance of qCMOS over conventional CCD technology is illustrated with simulations of short exposure time images of a star observed with a 1~m class ground based telescope.

\section{Computer Simulations comparing fast CMOS versus CCD}
\label{sec:simulations}

\begin{figure}[h!]
  \centering
\includegraphics[width=1.0\textwidth,bb=0 350 600 820,clip]{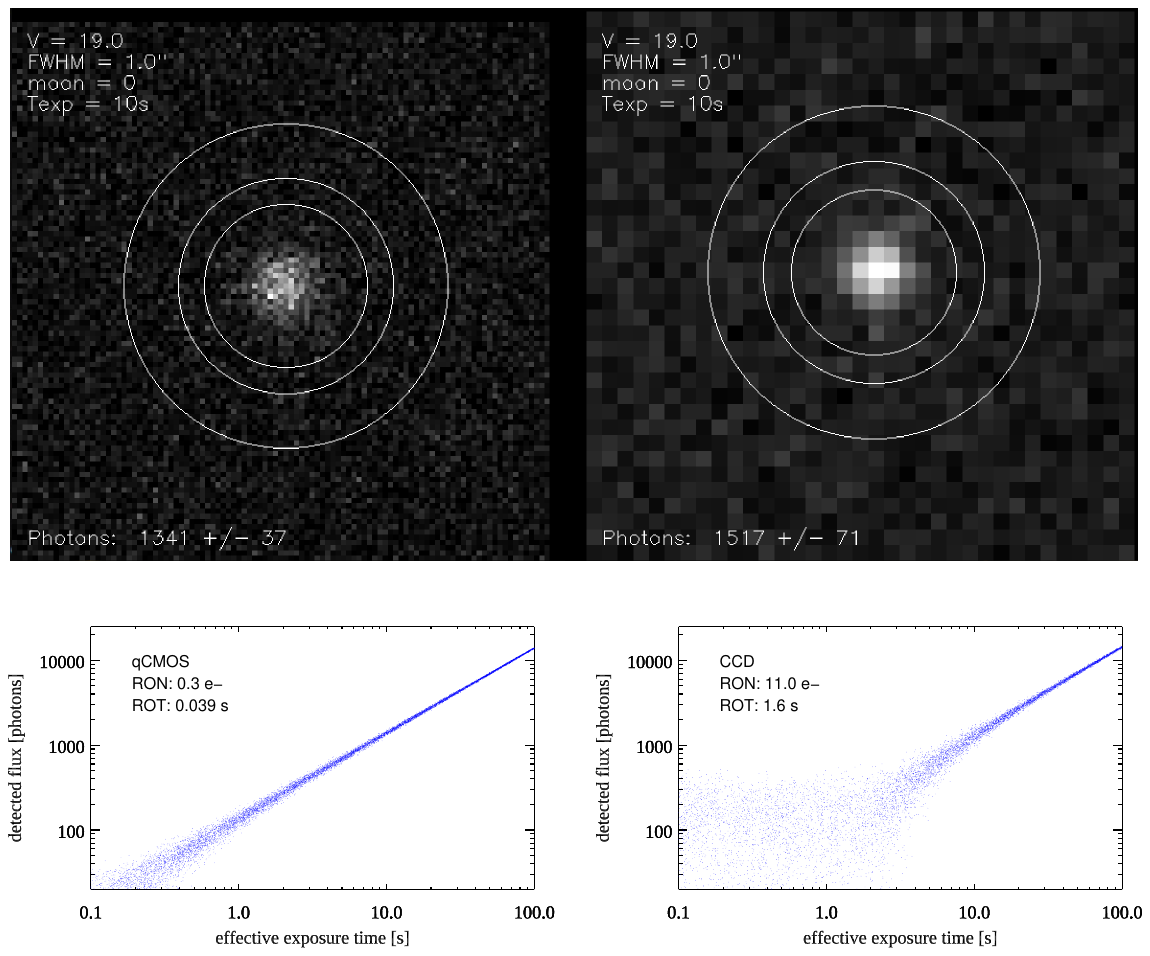}
  \caption{\small Simulation of the image of a star (V=19) showing results for qCMOS sensor (left) and CCD (right). In the example images on top, obtained for 10~s shutter open time for both sensors, the signal-to-noise ratio (SNR) is measured as 36 for the qCMOS, and 21 for the CCD. The scatter plots on the bottom present measured fluxes versus time series {\em effective integration time} for a total of 10\,000 realizations, illustrating how the CCD SNR rapidly degrades as the ROT becomes comparable to the time step.} 
 \label{fig:figure1}
\end{figure}

In order to achieve a meaningful performance comparison between the  two types of sensors, two camera systems that are available on the market were taken as reference and used with parameters listed in their data sheets: the Hamamatsu ORCA-Quest2 camera for the $4096\times2304$ / 4.6~$\mu$m pixel qCMOS, and the Andor iKon-L for the $2048\times2048$ / 13.5~$\mu$m pixel e2v CCD. The simulation was coded in IDL to create images of a star in a window of approximately $10\times10$~arcsec$^2$ at the sampling given by the qCMOS and CCD sensors, respectively, for a telescope of 1.23~m aperture, focal length of 9.8~m, and 0.58~m central obstruction. The flux for a star of given magnitude in a particular filter was computed from tabulated values for filters U, B, V, R, I, u, g, r, z, and multiplied with correction factors for the light collecting area of the telescope, and for atmospheric extinction and quantum efficiency at the respective wavelengths. The point-spread function was modeled with a Gaussian, scaled to the total stellar flux in units of photon counts for a given exposure time and the FWHM of atmospheric seeing, and then sampled at the  plate scale mentioned above. Likewise, the sky background was estimated from tabulated sky surface brightness values at the wavelength of a given filter for moon phases between new and full moon. The uncertainty of resulting photon counts in each pixel was modeled with Poissonian noise for counts from the star and sky, and Gaussian noise for the read noise (RON) of the camera.  The RON is quoted as 0.3~e$^-$ with readout time (ROT) 0.039~s for the ORCA-Quest2, and 11~e$^-$ with 1.6~s for the iKon-L, respectively. Finally, the stellar flux and its uncertainty was measured with DAOPHOT aperture photometry as indicated by the concentric circles in Fig.~1 for the aperture and the sky annulus around the centroid of the star. The measured stellar flux and its uncertainty is printed in the lower left corner of the plot. The two images represent exposures of the same shutter open time of 10~s for a star of V=19.0 at new moon. 

It is important to emphasize the difference between shutter open time for a {\em single} exposure and the effective exposure time for {\em time series exposures}, where the detector ROT must be subtracted from the time step between two exposures, thus clearly reducing the CCD efficiency when the time step is comparable to the ROT. Although the more scattered visual appearance of the qCMOS image may intuitively suggest a better result for the CCD in the 10~s shutter open example, the measured flux error shows the opposite. The scatter plots for a total of 10\,000 simulations demonstrate that the qCMOS has a clear advantage over the CCD at any exposure time, and that the SNR for the CCD dramatically breaks down below 10~s time series exposures.

\clearpage

\section{Future Steps and Developments}
\label{sec:conclusions}
The emerging Institute for Technology (ITE) at the upcoming German Center for Astrophysics (DZA) in Görlitz, Germany, is currently setting up a detector lab to characterize and evaluate two ORCA-Quest2 cameras that were delivered recently for applications as described above. After the formal foundation of DZA as a legal entity, envisaged for 2025, DZA is planning to deploy the cameras at two remote 1~m class telescopes to validate the concept on sky, including the option to directly measure parallaxes of NEOs by triangulation. This research note is intended to alert the community of the capabilities of the qCMOS technology, and to potentially spawn collaborations with other interested parties.

\section{Acknowledgements}
\label{sec:acknowledgements}
DZA research on CMOS sensors for astronomy is supported by BMBF under grant 03WSP1745.

\bibliography{qCMOS-pub}

\end{document}